\documentclass[aps,pre,longbibliography,twocolumn,superscriptaddress,showpacs]{revtex4-2}

\usepackage{CJK}
\usepackage{amsmath}
\usepackage{graphicx}
\usepackage{dcolumn}
\usepackage{bm}
\usepackage[normalem]{ulem}
\usepackage{xcolor}
\usepackage{amsmath}
\usepackage{epstopdf}
\usepackage{verbatim}

\newcommand{\br}{\mathbf{r}}

\newcommand{\bu}{\mathbf{u}}

\newcommand{\hx}{\hat{x}}
\newcommand{\hy}{\hat{y}}

\newcommand{\beq}{\begin{equation}}
\newcommand{\eeq}{\end{equation}}
\newcommand{\beqn}{\begin{eqnarray}}
\newcommand{\eeqn}{\end{eqnarray}}
\newcommand{\beqa}{\begin{align}}
\newcommand{\eeqa}{\end{align}}

\newcommand{\dd}{{\rm d}}

\newcommand{\bee}{{\bf e}}
\newcommand{\eq}{Eq.\ }

\begin{document}

	\title{Modeling growing confluent tissues using a lattice Boltzmann method: interface stability and fluctuations}
	\author{Andrew Killeen}
	\affiliation{Department of Bioengineering, Imperial College London, South Kensington Campus, London SW7 2AZ, U.K.}
	\author{Benjamin Partridge}
	\affiliation{Department of Bioengineering, Imperial College London, South Kensington Campus, London SW7 2AZ, U.K.}
	\author{Thibault Bertrand}
	\email{t.bertrand@imperial.ac.uk}
	\affiliation{Department of Mathematics, Imperial College London, South Kensington Campus, London SW7 2AZ, U.K.}
	\author{Chiu Fan Lee}
	\email{c.lee@imperial.ac.uk}
	\affiliation{Department of Bioengineering, Imperial College London, South Kensington Campus, London SW7 2AZ, U.K.}
	
\begin{abstract}	
Tissue growth underpins a wide array of biological and developmental processes, and numerical modeling of growing systems has been shown to be a useful tool for understanding these processes. However, the phenomena that can be captured are often limited by the size of systems that can be modeled. Here, we address this limitation by introducing a Lattice-Boltzmann method (LBM) for a growing system that is able to efficiently model hydrodynamic length-scales. The model incorporates a novel approach to describing the growing front of a tissue, which we use to investigate the dynamics of the interface of growing model tissues. We find that the interface grows with scaling in agreement with the Kardar-Parisi-Zhang (KPZ) universality class when growth in the system is bulk driven. Interestingly, we also find the emergence of a previously unreported hydrodynamic instability when proliferation is restricted to the tissue edge. We then develop an analytical theory to show that the instability arises due to a coupling between the number of cells actively proliferating and the position of the interface.
\end{abstract}
\maketitle
\section{Introduction}
Many biological processes, from cancer metastasis to morphogenesis, rely on the integration of cell proliferation and collective cell movement. While cell proliferation is known to be regulated by the mechanical properties of the tissue \cite{Mammoto2010}, it is also becoming increasingly clear that proliferation alters the properties, and consequently the dynamics, of the tissue in return \cite{Doostmohammadi2015,Ranft2010,Etournay2015}.  Whereas the forces generated during cell division events are well understood \cite{Rossen2014}, our understanding of how these affect dynamics at the collective level is comparatively more limited \cite{Hakim2017}. This is due to the difficulty in determining how cellular scale processes, such as cell division, manifest themselves as macroscopic dynamics \cite{Anderson1972}.

Numerical models offer a fruitful avenue for exploring growing and collectively migrating biological systems \cite{Marchetti2013}. Cell based simulations have shown how cell division affects the structure of, and fluidizes, epithelial tissues \cite{Devany2021,Malmi-Kakkada2018,Czajkowski2019,MatozFernandez2017}. They have also delineated the effect of cell division on different macroscopic dynamics, such as on coherent angular motion in morphogenesis \cite{Siedlik2017} and how the interplay of mechanical stresses and cell proliferation can drive fronts of growing cells \cite{Li2021}. However, due to the increase in computational complexity as the number of cells increase, cell based models are limited in the length-scale of system they can model, leading to a need for methods that can describe these growing systems in the hydrodynamic limit. 

Continuum models have been employed to study growing bacterial colonies in two \cite{You2018,DellArciprete2018} and three dimensions  \cite{Pearce2019}, while multiple studies have described growing biological systems in 2D using a hybrid lattice-Boltzmann method (LBM) \cite{Doostmohammadi2016,Kempf2019,Doostmohammadi2015}. LBMs are an incredibly efficient means of coarse-grained modeling. Well established as an efficient means of simulating passive fluids, more recently active nematic \cite{Marenduzzo2007} and polar \cite{Nesbitt2021} systems have been described using LBMs. Previous LBM studies of active systems have either used periodic domains in which the entire system is active, or have described the boundary of the active material using a phase separating system with a phase field \cite{Doostmohammadi2015}. These approaches are valid if the phenomena under study relate to behavior in the bulk or if the system is not completely phase separated, i.e. the `vapor' phase has a non-vanishing, albeit low, density. However, they are not appropriate in systems with well-defined boundaries and areas completely devoid of cells, such as expanding tissue layers or densely packed bacterial biofilms. 

In such systems, properly capturing the interface is critical to an accurate description of important physiological processes. As such, in recent years, numerous studies have sought to characterize the dynamics of the interface. This has been done by examining the stability of growing tissue fronts under different conditions \cite{Alert2019,Trenado2021,Nesbitt2017,Zimmermann2014} or by ascertaining the scaling behavior of the interface roughness to determine the universality class to which the growth process belongs \cite{Mazarei2022,Azimzade2019,Huergo2010,Lee2017,Besse2022,Patch2018}. Importantly, these studies led to conflicting results, leaving the question of the dynamics of growing tissue fronts and their stability unsettled. The observed disagreements are likely due to the difficulty in simulating large enough systems; to address this outstanding issue, we here develop a methodology able to overcome these computational challenges.

%%%%%%%%%%%%%%%%%%%%
\begin{figure*}[t!]
    \begin{center}
        \includegraphics[width=172mm]{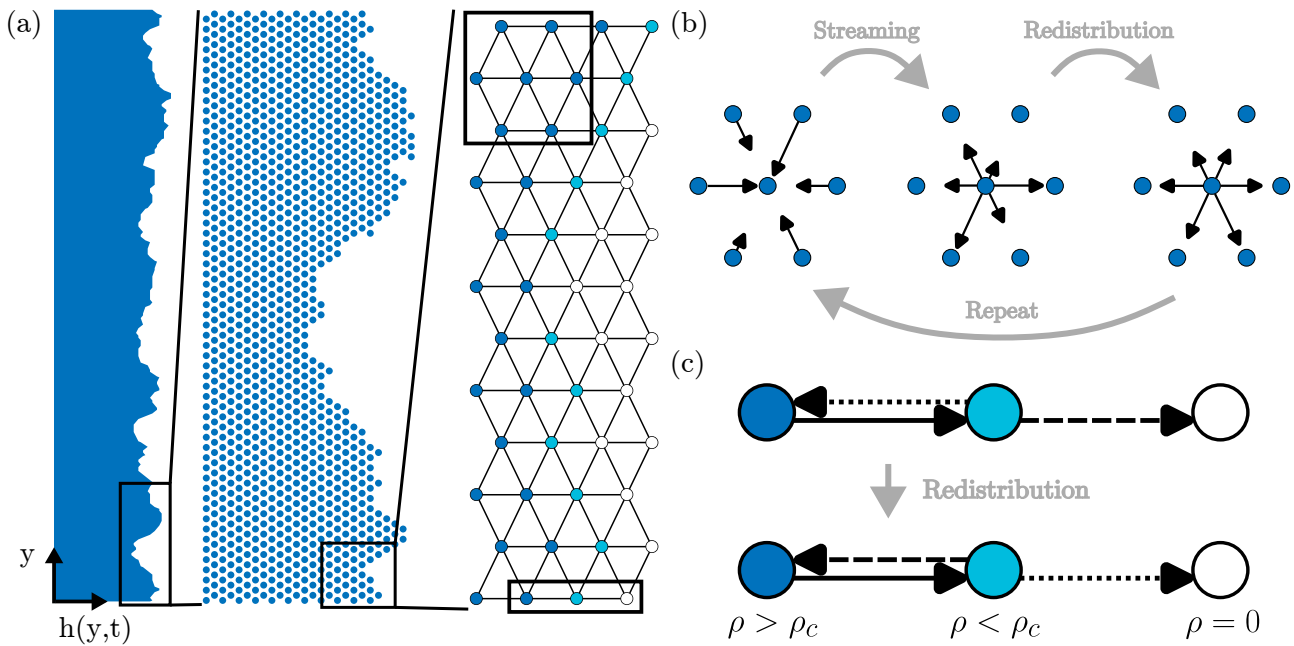}
    \end{center}
    \caption{A LBM for a growing biological system. (a) We discretize our system (left) onto a hexagonal lattice (middle), with each lattice point representing a number of cells. There is a growing front beyond which the distribution function $f_i$ is zero in all lattice directions, (white sites in right). (b) Each time-step is subdivided into a streaming step and redistribution step. In the streaming step, $f_i$ is advanced by one time-step to get an intermediate distribution $f_i^*$. A steady-state distribution $f_i^{ss}$ is then calculated using $f_i^*$ and $f_i$ is redistributed such that it relaxes towards $f_i^{ss}$. (c) The growing periphery is enforced using a `bounce-back' condition during the redistribution step. When the density at a site is below a critical density $\rho_c$, instead of relaxing $f_i$ towards the steady state distribution $f_i^{ss}$, the directions of the distribution function are reversed such that the mass that would next stream to an empty site is now bounced back in the opposite direction (dashed arrow). The mass just streamed from an empty site, which would be zero, is then set to stream back out instead (dotted arrow).}
    \label{fig1}
\end{figure*}
%%%%%%%%%%%%%%%%%%%%%%%%

In this article, we address these issues by introducing a novel LBM for growing biological tissues capable of describing faithfully the dynamics of growing fronts.  Crucially, our method for modeling the interface of our tissue ensures there is no low density vapor phase and our tissue has a sharp edge. Using this model, we study the dynamics and stability of a growth front. Specifically we find that the interface of a tissue growing due to bulk driven cell proliferation pertains to the Kardar-Parisi-Zhang (KPZ) universality class. We then consider tissue growth with a density dependent proliferation regime where proliferation occurs around the growing front, where we again find KPZ scaling in the interface width for small system sizes. However, we also find a previously unreported mechanical instability at larger system sizes, which we explain using a linear stability analysis. These findings demonstrate the efficacy of using this LBM, as its mesoscopic scale and efficiency made it possible to simulate large enough systems to observe this instability and allow us to clearly exhibit KPZ scaling. 

The paper is organized as follows. In section II, we introduce our novel LBM for a growing system with a fluctuating interface. In section III, we characterize the scaling behavior of the interface using a bulk driven growth regime. In section IV, we then study interface fluctuations using a density dependent growth regime, before examining the emergent instability and how it forms. We then discuss our model and conclusions in section V.

\section{Lattice Boltzmann method}
The key development of our LBM is the inclusion of cell proliferation and a growing interface that properly captures the behaviour of tissue interfaces by having a sharp boundary, beyond which is completely devoid of any mass. However, for context, we first give an overview of the LBM.

As the starting point for our present model we take a recently developed LBM for a dry active fluid system \cite{Nesbitt2021}. The efficiency of the LBM stems from solving, instead of the hydrodynamic equations of motion (EOM), a simplified system that obeys the same hydrodynamic symmetries as the real system, leading to identical behavior in the hydrodynamic limit \cite{Chen1992,Krueger2016}. This is done by calculating the dynamics of a discretized distribution function $f_i(t, \br)$, which represents the distribution of mass at time $t$ and position $\br$, where $i$ corresponds to directions on the lattice on which our system is discretized. We use a two-dimensional triangular lattice [Fig.\,\ref{fig1}(a)], termed D2Q7 in standard LBM notation, meaning mass moves along lattice vectors $\bee_i = \cos{[(i-1)\pi/6]}\mathbf{\hx}+\sin{[(i-1)\pi/6]}\mathbf{\hy}$ for $i \in \{1,2,...6\}$ and $\bee_0 = 0$, corresponding to mass which is at rest. The lattice speed $c$ is the ratio of our grid spacing $\Delta x$ to time-step $\Delta t$, which we choose to be 1. The hydrodynamic variables of interest, cell density $\rho$ and velocity $\bu$ can then be calculated from $f_i$ using
\beq
\rho(t,\br) = \sum^6_{i=0} f_i(t, \br) \, , \ \bu(t,\br) = \frac{\sum^6_{i=0} f_i(t, \br)\,c\bee_i}{\rho(t,\br)} \ .
\eeq

We then evolve $f_i$ according to $f_i(t+\Delta t,\br + c\bee_i\Delta t) = f_i(t, \br) \, + \, \Omega_i$, where $\Omega_i$ is a collision operator that ensures matter in our system interacts whiles obeying the same symmetries as the system we wish to model. While there are many choices for $\Omega_i$, we use the Bhatnagar-Gross-Krook collision operator \cite{Bhatnagar1954} with additional fluctuations $\eta_i$, meaning $f_i$ evolves according to:
\beq
f_i(t+\Delta t,\br + \bee_i) = f_i(t, \br) - \frac{1}{\tau} \left[ f_i(t, \br) - f_i^{SS}(t, \br)\right ] + \eta_i(t, \br) \ ,
\label{eq:boltzmann}
\eeq
where we use $\br+ \bee_i$ for brevity, as $\br+ \bee_i=\br+\Delta x\bee_i=\br + c\bee_i\Delta t$. Our relaxation parameter is $\tau$, $\eta_i$ are random fluctuations and $f_i^{SS}$ is a steady state distribution which respects the same symmetries as our system, meaning our system obeys the correct symmetries as it relaxes towards $f_i^{SS}$. While `steady state' implies a quantity that does not change, $f_i^{SS}$ refers to the distribution function the system would have were it in steady state given the current density and velocity fields, and so it changes as these quantities change. We define $f_i^{SS}$ as
\beq
f_i^{SS}=w_i\rho\left(1+4\frac{\bee_i\cdot\bu^*}{c}+8\frac{(\bee_i\cdot\bu^*)^2}{c^2}-2\frac{|\bu^*|^2}{c^2}\right) \ ,
\label{eq:f_ss}
\eeq
where $w_i$ are lattice direction weights, with $w_0=1/2$ and $w_{i\neq0}=1/12$, and $\bu^*$ is our steady-state velocity that depends on the system we are modeling. We define $f_i^{SS}$ in this way  because, if $\bu^* = \bu$, \eq (\ref{eq:f_ss}) would be the equilibrium distribution used to model passive fluids in 2D with a triangular lattice, as it conserves mass and momentum \cite{Krueger2016}. However, as active systems do not require the conservation of momentum, we can choose $\bu^*$ to be any function of $\bu$ and $\rho$ that respects the symmetries of the system of interest. As our focus here is to study the effects of proliferation and not motility, here we choose $\bu^* = (1 - \mu)\bu$, where $\mu$ encodes dissipation arising from friction with the substrate, although different forms of $\bu^*$ can be used to model active self-propulsion \cite{Nesbitt2021}. As we have defined it, \eq (\ref{eq:boltzmann}) always conserves mass regardless of our choice of $\bu^*$, which is obviously not the case in a growing system. However, choosing $f_i^{SS}$ so as to conserve mass allows easier control over precisely how mass is added to our system and so how cell proliferation is modeled.

The fluctuations $\eta_i$ in \eq (\ref{eq:boltzmann}) are defined as
\beq
\eta_i(t, \br) = \tilde{\eta}_i(t, \br) - \frac{1}{7}\sum_{i=0}^6\tilde{\eta}_i(t, \br) \ ,
\eeq
where $\tilde{\eta}_i$ is an uncorrelated random variable that is uniformly distributed between $[-\sigma,\sigma]$. This form of noise is chosen so as to conserve mass, for reasons discussed previously.

We evolve \eq (\ref{eq:boltzmann}) in two steps: a streaming step and a redistribution step [Fig.\,\ref{fig1}(b)]. In the streaming step $f_i$ is evolved by one time-step to give an intermediate distribution $f_i^*(t+\Delta t,\br + \bee_i) = f_i$. In the redistribution step, $f_i^{SS}$ is then calculated at each point by calculating the hydrodynamic variables based on $f_i^*$. We then relax $f_i^*$ towards $f_i^{SS}$ using \eq (\ref{eq:boltzmann}) by replacing $f_i(t, \br)$ with  $f_i^*(t+\Delta t,\br + \bee_i)$ on the right hand side. Fluctuations $\eta_i$ are then added to give the final distribution at the next time-step $f_i(t+\Delta t,\br + \bee_i)$. 

From this model, we incorporate both cell proliferation and a growing tissue boundary. To model cell proliferation, inbetween the streaming and redistribution steps, we add mass at a chosen a site by increasing $f_i$ of two opposite lattice vectors, for example $f_1$ and $f_4$, by $\rho_{\rm cell}/2$. Here $\rho_{\rm cell}$ corresponds to the mass of one cell. The direction in which mass is injected is chosen at random and replicates the extensile nematic nature of cell division \cite{Doostmohammadi2015}. We determine the number of sites at which we inject mass by setting growth to be at a constant rate $g$. The number of sites to be randomly selected is then $n_s = g\,m_{\rm tot}/\rho_{\rm cell}$, where $m_{\rm tot}$ is the total cell mass, found by summing $\rho$ over all lattice points. Proliferation sites are then selected, with replacement, at random until enough mass has been added to the system. Time or position dependant proliferation rates can then be implemented depending on the probability distribution from which proliferation sites are selected. 

%%%%%%%%%%%%%%%%%%%%
\begin{figure*}[t!]
    \begin{center}
        \includegraphics[width=172mm]{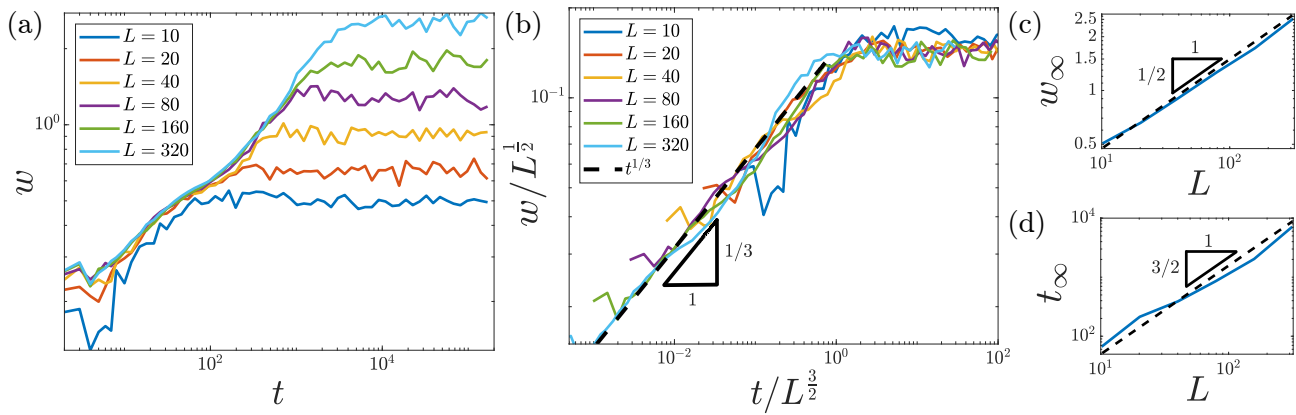}
    \end{center}
    \caption{KPZ scaling. (a) Time evolution of the interface width $w$ for systems of different length $L$. (b) Curves collapse when rescaled by KPZ exponents. The dashed line showing scaling $t^{1/3}$ highlights the corresponding KPZ growth exponent. KPZ scaling of (c) $w_{\infty}$ and (d) $t_{\infty}$ with $L$. The dashed line is a guide showing KPZ scaling for (c) the roughness exponent $\alpha$, $w_\infty \sim L^{1/2}$ and (d) the dynamic exponent $z$, $t_\infty \sim L^{3/2}$.}
    \label{fig2}
\end{figure*}
%%%%%%%%%%%%%%%%%%%%%%%%

Here, we look at two growth regimes. Initially, we select proliferation sites from a uniform distribution, meaning proliferation in the active portion of our tissue is equally likely in any occupied site in the bulk or at the interface, meaning the majority occurs in the bulk. We use the phrase `active portion' because, as the tissue grows and the interface advances, we advance the rear wall of the model with it. This greatly improves efficiency and reflects the fact that biological tissues typically have a section towards the front that is actively proliferating, with cells far from the boundary becoming quiescent. Secondly, we implement a more biologically relevant growth regime where the local proliferation rate is dependent on the local density in the system, which concentrates proliferation to the interfacial region. This is motivated by increased cell density causing increases in tissue pressure away from the tissue boundary, \cite{Barton2017}, which can inhibit cell proliferation \cite{Montel2012}. Now, when a lattice site is randomly selected, cell division occurs with probability $p_{\rm div}(\br)$, which decays linearly with the local density according to 
\beq
p_{\rm div}(\br) = 
    \begin{cases}
    1-\rho(\br)/\rho_0\ , & \text{if}\ \rho(\br)>\rho_c \\
      0\, & \text{otherwise}
    \end{cases}
\label{eq:var_prolif}
\eeq
where $\rho_0$ the critical density above which proliferation ceases. This regime models a scenario where crowding suppresses proliferation due to increased compressive stresses, such as in epithelial layers \cite{Puliafito2012}, meaning the bulk of the tissue becomes quiescent.

\subsection*{Moving growth front: a two-step bounce-back method}

While growth in the form of mass injection can be readily implemented within the LBM framework, a key difficulty in applying it to a growing system is how the interfacial dynamics should be captured. Here, our key innovation is to use a two step thresholding method to accomplish this task. Specifically, our method ensures that the `vapor' phase is completely devoid of any mass, which also distinguishes our method from existing LBMs applied to phase separating systems. This allows the proper modeling of the tissue layer's boundaries using a LBM for the first time. We achieve this by developing a type of freely moving bounce-back method. In the redistribution step, if the density at a given site is below a threshold value $\rho_c$, instead of relaxing towards $f_i^{SS}$, the directions of $f_i^*$ are reversed such that $f^*_i=f^*_j$ where $\bee_j$ is the reverse direction of $\bee_i \,(1\leftrightarrow 4, 2\leftrightarrow 5, 3\leftrightarrow 6)$. This means the mass just streamed to a given node in the stream step is reflected such that it now travels back in the direction it came from [Fig\, \ref{fig1}(c)]. This ensures that any mass that would be streamed `out' of the system in the next time-step is rebounded back in and the lattice site it would be streamed to remains empty. This ensures that at the edge of the tissue there is one lattice site with $\rho<\rho_c$ and beyond this the system is devoid of mass. This mimics a surface tension like force and allows our LBM to easily model sharp interfaces such as an epithelial tissue edge. 

\section{Bulk driven growth}
To demonstrate the efficacy of our model, and the utility of our method for capturing the dynamics of the interface, we use our model to study the dynamics of the tissue boundary, or interface, initially in a regime where proliferation is taken to happen uniformly across the active portion of the tissue, meaning the majority of it occurs in the bulk. Growing interfaces are characterized by calculating how the roughness of the interface --- or interface width --- scales with space and time. Calculating the scaling behavior of the interface width allows one to determine critical exponents and hence the universality class to which the system belongs  \cite{Barabasi1995,Family1985}. As we are dealing with a nonequilibrium, growing system, we would generically expect that it belongs to the KPZ universality class  \cite{Barabasi1995}. 

Interface growth in biological systems has been studied experimentally \cite{Huergo2010,Bru1998,Bru2003,Galeano2003} and numerically \cite{Block2007,Santalla2018,Mazarei2022,Azimzade2019} and, although KPZ scaling has been observed \cite{Huergo2010,Block2007,Santalla2018,Mazarei2022}, there is some debate as to what the proper scaling is. Scaling in agreement with the molecular beam epitaxy universality class has also been reported \cite{Bru1998,Bru2003}, along with scaling inconsistent with either class  \cite{Galeano2003,Vicsek1990}. This debate often stems from the difficulty in simulating large enough systems or subtleties in the models used, such as heterogeneity in the surrounding environment \cite{Azimzade2019}. It can also arise from the difficulty of using experimental data to perform the scaling analysis \cite{Buceta2005}. Using a model as generic and efficient as our LBM allows for insight into how cell proliferation manifests itself in interface growth without these subtleties.

%%%%%%%%%%%%%%%%%%%%
\begin{figure*}[t!]
    \begin{center}
        \includegraphics[width=172mm]{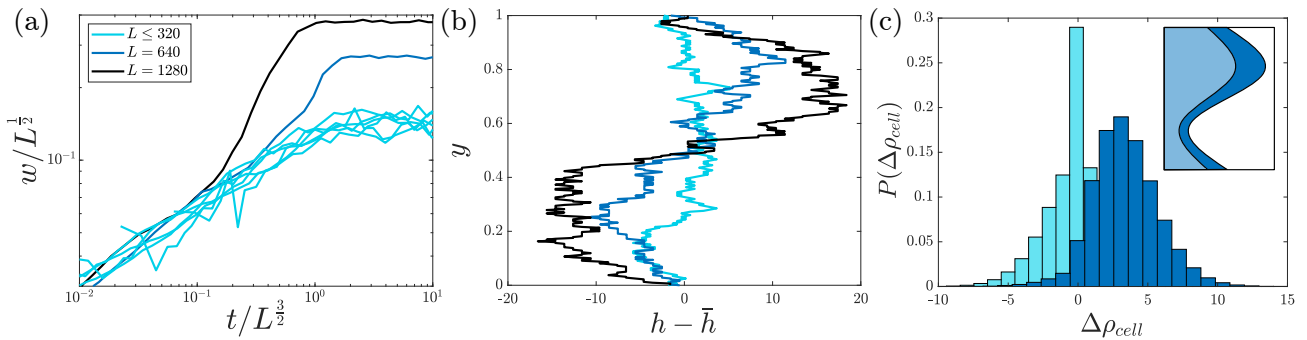}
    \end{center}
    \caption{A system spanning instability causes divergence from KPZ scaling. (a) Interface growth curves for systems with $L < 320$ collapse when rescaled by KPZ exponents, but diverge for $L > 320$. (b) Examples of steady-state profiles of the interface position $h$ shifted by the average interface position $\bar{h}$ for systems with $L=320$ (light blue), $L=640$ (dark blue) and $L=1280$ (black). (c) Probability distributions for the difference between the number of cell divisions $\Delta\rho_{\rm cell}$ around $h_{\rm max}$ and $h_{\rm min}$ for $L=40$ (light blue) and $L=640$ (dark blue). (inset) Schematic of the proposed instability mechanism, proliferative regions are shown in dark blue. The more advanced sections of the interface are more proliferative.}
    \label{fig3}
\end{figure*}
%%%%%%%%%%%%%%%%%%%%%%%%

To examine how the interface evolves, we calculate the interface width $w$, defined as the standard deviation of the interface height $h(y,t)$
\beq
w(t) =\sqrt{ \frac{1}{L} \int^L_0\Bigl[h(y,t)-\bar{h}(t)\Bigl]^2\dd y} \ ,
\eeq
where $L$ is the width of the domain. How the interface scales in time and space can be completely described by two scaling exponents $\alpha$ and $\beta$. The growth exponent $\beta$ describes how $w$ grows before it reaches its steady-state, saturation value $w_{\infty}$ at time $t_{\infty}$, that is $w \sim t^\beta \; (t\ll t_{\infty})$. The roughness exponent $\alpha$ describes how $w_{\infty}$ scales with the system size $L$, $w_{sat} \sim L^{\alpha} \; (t\gg t_{\infty})$. The time for the system to reach steady-state then scales according to the dynamic exponent $z = \alpha / \beta$, meaning $t_{\infty} \sim L^z$. For KPZ scaling one would expect $\alpha=1/2$ and $\beta=1/3$, leading to $z = 3/2$.

To investigate this, we implement our LBM, with proliferation uniform throughout the non-quiescent region. We implement it on rectangular domains of different widths $L$, with periodic boundary conditions at the top and bottom boundaries, and a bounce-back condition at the rear wall. A full description of the implementation, along with a complete list of parameter values used, can be found in Appendix A. 

Fig.\,\ref{fig2}(a) shows the time evolution of $w$ for different system sizes, showing, as anticipated, that each systems interface grows at the same rate but that larger systems permit rougher interfaces. Upon rescaling $w$ and $t$ by the appropriate KPZ exponents, $L^{1/2}$ and $L^{3/2}$ respectively, we find a very good curve collapse, indicating the system is exhibiting KPZ scaling [Fig.\,\ref{fig2}(b)]. This is underlined in Fig.\,\ref{fig2}(c) and (d), which show the appropriate scaling of $w_{\infty}$ and $t_{\infty}$ with $L$. 

\section{Density dependent growth}
\subsection*{Interface growth}
We now implement a more realistic scenario, where proliferation is concentrated towards the boundary of the system. We do this by implementing a density dependent growth regime following Eq. (\ref{eq:var_prolif}), keeping other parameters the same. Upon studying the growth of the interface in this regime, we again see KPZ scaling for systems with $L<640$, however, for larger system sizes KPZ scaling is not observed and the interface width appears to diverge [Fig.\,\ref{fig3}(a)]. 

To investigate the cause of this divergence, we plot the steady-state profile of the boundary for systems of different sizes  [Fig.\,\ref{fig3}(b)]. Surprisingly, upon doing this we see that, for sufficiently large $L$, the interface is subject to a system-spanning instability, at the largest wavelength permitted by the system. To gain an insight into the cause of the instability, we studied proliferation in the tissue in the area around the most advanced ($h_{\rm max}$) and least advanced ($h_{\rm min}$) positions of the interface. We define these areas as rectangular regions extending 10 lattice sites above and below $h_{\rm max}$ or $h_{\rm min}$, and 20 lattice sites into the bulk. We find that, for larger system sizes, there were more proliferation events around $h_{\rm max}$ than $h_{\rm min}$, although this pronounced bias was not seen for smaller system sizes  [Fig.\,\ref{fig3}(c)]. This led us to hypothesize that a local increase in the proliferation rate around protrusions to the interface could lead to an increase in the local interface velocity, thus driving the instability  [Fig.\,\ref{fig3}(c) inset]. To understand the mechanism driving the instability further we now develop a minimal model of interface growth in our system.

\subsection*{Linear stability analysis}
\begin{figure}[t!]
    \begin{center}
        \includegraphics[width=80mm]{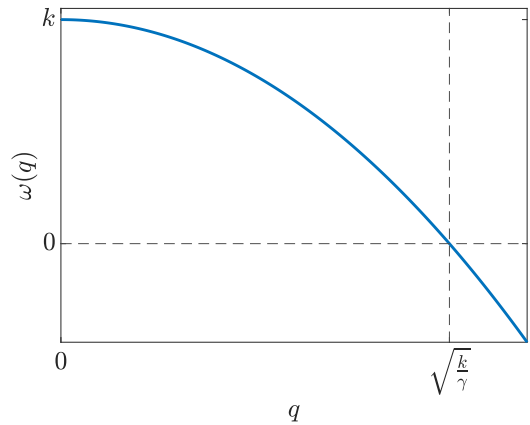}
    \end{center}
    \caption{Growth rate of perturbations of different wave numbers. Perturbations only become unstable at wave numbers $q< \sqrt{k/\gamma}$, corresponding to lengths $L > \sqrt{\gamma/k}$. Beyond this threshold, the fastest growing mode is always the longest wavelength permitted by the system. }
    \label{fig4}
\end{figure}

The mechanism driving the instability is unclear as, while instabilities in active systems arising from motility have been previously reported \cite{Alert2019,Trenado2021,Nesbitt2017,Zimmermann2014}, instabilities arising from cell proliferation have been less well studied. Cell proliferation has been found to induce an instability at a boundary between two tissues \cite{Basan2011,Buscher2020b}, or when a tissue is growing into another viscous medium \cite{Bogdan2018,Williamson2018}, but an instability arising from a purely proliferative system expanding into a void has not been reported in the literature. To understand the mechanism at play, we write down an equation of motion for the interface position $h$.  Our numerical results suggest that for the instability to occur we require the growth rate of the interface to increase as we move from $h_{\rm min}$ to $h_{\rm max}$, and so be proportional to $h$. This can be understood by considering what happens when the boundary locally advances. Any local increase in $h$ will necessarily have a lower density than the area preceding it and so a higher likelihood of cell division. Due to friction, and the limits on how quickly information can propagate in the system, the effect of these changes in density can only propagate back into the bulk at a finite speed. Consequently, if the timescale for cell proliferation is shorter than that of this propagation, areas where the interface is more advanced grow faster. Along with this, surface tension, coming from the bounceback condition in our LBM, also clearly has an effect on interface dynamics. The dynamics of $h$ are thus, to lowest order, governed by 
\beq
\partial_th(y,t) = kh(y,t) + \gamma\partial^2_yh(y,t) \ ,
\label{eq:h_eqn}
\eeq
where $k$ is the growth rate and $\gamma$ the surface tension. We note that this equation is far more general than our LBM and applies to any system in which the growth rate of the interface depends on its position. To probe the stability of this system, we perform linear stability analysis on Eq. (\ref{eq:h_eqn}). We add a small amplitude perturbation of the form $\delta h=h_0e^{\omega t+iqy}$, where $|\delta h|\ll1$ and $\omega$ describes the growth rate of each wave number $q$, to a flat interface in a frame of reference comoving with the mean interface height. Doing so yields the dispersion relation
\beq
\omega(q) = k - \gamma q^2 \ ,
\label{eq:dispersion}
\eeq
which can be seen plotted in Fig.\,\ref{fig4}. From Fig.\,\ref{fig4} it is clear that the fastest growing mode will always be the largest one permitted by the system, the system size, and that the growth rate can be positive if wave numbers less than $\sqrt{k/\gamma}$ are permitted, corresponding to system sizes $L>\sqrt{\gamma/k}$. This is why the instability is only seen at sufficiently large system sizes, as the system only becomes unstable at a critical length $L_c = \sqrt{\gamma/k}$. However, we note that $L_c$ is dependent on system parameters and so may not be very large if the growth rate is sufficiently fast or the surface tension weak.

%%%%%%%%%%%%%%%%%%%%
%%%%%%%%%%%%%%%%%%%%%%%%

To demonstrate that this is indeed the instability mechanism, we implement a different growth regime while keeping the overall growth rate constant. We restrict growth to being in a specified section of lattice sites in the bulk, far from the interface. We do this in a system of size $L=640$, where the instability was previously observed. While the biological plausibility of this scenario is debatable, introducing this type of growth removes the purported instability mechanism as the width of proliferating region is constant across the entire domain length $L$ and so independent of the interface position. As can be seen in Appendix B, introducing this growth regime eliminates the instability. These results suggest that {\it any} growing system, where the number of constituents that are able to proliferate increases locally where the interface advances, could be susceptible to this system spanning instability. 

\section{Discussion \& outlook}
We present a novel LBM for modeling growing, active systems. Our bounce-back method for the interfacial dynamics ensures well-defined, freely moving boundaries that, for the first time, allows for the proper physics of growing tissues to be captured using a LBM. Using this model, we demonstrate that the growth of the boundary driven purely by proliferation displays KPZ-like scaling, but also displays an instability in physiologically relevant proliferation regimes where proliferation is concentrated at the boundary. We formulate an analytical theory to demonstrate this instability arises due to a proliferation rate dependant on the position of the interface. This theory is far more general than the particular model in which we observed the instability, and asserts that any system where the local growth rate is dependant on the interface position will be susceptible to this instability.

The results presented here are for a single value of the friction coefficient $\mu$. However, we note that for very high values of $\mu$ the system appears to undergo a roughening transition where the steady-state interface width becomes independent of the system size $L$. Roughening transitions are known to occur when fluctuations on the system are sufficiently suppressed \cite{Barabasi1995} and an interesting avenue of future work would be to investigate this transition, and the impact of friction on the system more generally, in more detail.

Also, here the only activity in the system is due to cell proliferation, focusing our study on systems where the dynamics are dominated by growth processes. However, there are many biological contexts where cell motility, which can be included in the model due to its flexibility \cite{Nesbitt2017}, plays an important role in the collective behavior of the system. Exploring the effect of motility, its interplay with cell division, and the impact this has on the dynamics of the boundary and the onset of instability, presents an interesting avenue for future work.

\begin{acknowledgments}
We acknowledge the High Throughput Computing service provided by Imperial College Research Computing Service. AK was supported by the EPSRC Centre for Doctoral Training in Fluid Dynamics Across Scales (Grant EP/L016230/1).
\end{acknowledgments}
\appendix

\renewcommand\thefigure{\thesection.\arabic{figure}}    
  
\section{LBM Implementation}
\setcounter{figure}{0}  
We initialise our system by setting the initial velocity field to 0 everywhere and generating an initial density distribution with a flat interface, centered around some initial average density $\rho_{\rm init}$, uniformly distributed on $[0.9 \rho_{\rm init}, 1.1\rho_{\rm init}]$. From these initial distributions we calculate an initial $f_i^{SS}$, which we set as our initial $f_i$.

The algorithm then proceeds in the following steps:
\begin{enumerate}
    \item Streaming step to generate intermediate distribution $f_i^*(t+\Delta t,\br + c\bee_i\Delta t) = f_i$.
    \item Calculate $\rho$ and $\bu$ from  $f_i^*$.
    \item Add mass from cell proliferation.
    \item Calculate $\bu^*$ and hence $f_i^{SS}$.
    \item Redistribution step:
	 \begin{itemize}
		\item If $\rho(\br)\,\geq\,\rho_c$:  redistribute $f_i \,=\, f_i^* - \left[\, f_i^*\, -\, f_i^{SS}\,\right ]\,/\,\tau$.
		\item If $\rho(\br)<\rho_c$:  set $f_i=f_j$ where $\bee_j$ is the reverse direction of $\bee_i \,(1\leftrightarrow 4, 2\leftrightarrow 5, 3\leftrightarrow 6)$.
	  \end{itemize}
    \item Add noise: $f_i\to f_i + \eta_i$, provided $\rho>\rho_c$.
    \item Correct any negative values of $f_i$ that arise from adding noise:
	  \begin{itemize}
		\item If $f_0<0$:  set $f_0=0$. This will change the density, so rescale each direction according to $f_i \to \rho f_i/\Sigma_j f_j$.
		\item If $f_i<0$ for $i>0$:  for $i>0$, set $f_j=f_j + |f_i|$ where $\bee_j$ is the reverse direction of $\bee_i \,(1\leftrightarrow 4, 2\leftrightarrow 5, 3\leftrightarrow 6)$ and set $f_i=0$.
	  \end{itemize}
\end{enumerate}

As our tissue grows we advance the rear wall such that it is always at least ${\rm max}(50,10w)$ lattice units behind $\bar{h}$. We use system parameters $c =1$, $\rho_c=0.05$,  $\tau=1$,   $\rho_{\rm init}=0.1$, $\sigma=0.01$,  $\mu=0.001$,  $g=0.001$ and $\rho_{\rm cell}=0.01$. For the density dependent growth regime we use $\rho_0=0.15$.

\section{Suppressing the instability}
\begin{figure*}[ht!]
    \begin{center}
        \includegraphics[width=172mm]{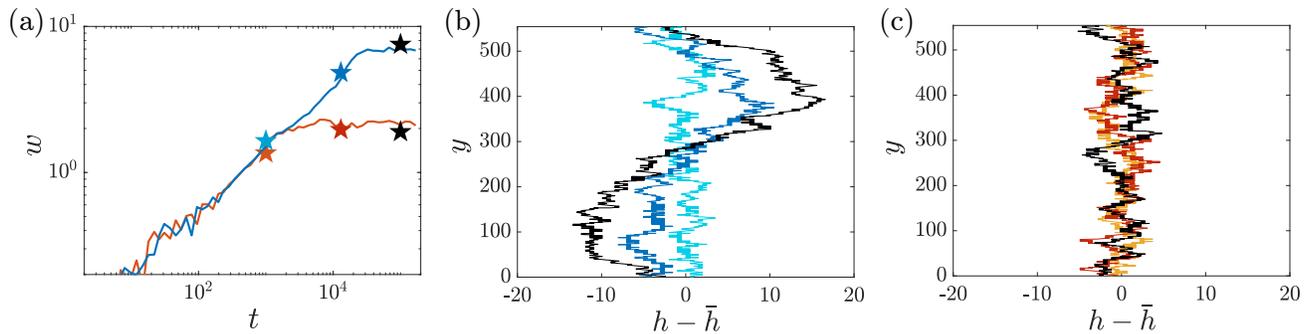}
    \end{center}
    \caption{Restricting proliferation to the bulk removes the instability. (a) Time evolution of $w$ when proliferation is (blue) density dependent and (orange) permitted only in the bulk. Stars indicate times at when profiles of the same colour are depicted in b and c. Time evolution of the interface position $h$ shifted by the average interface position $\bar{h}$ for a system of size $L=640$ with (a) density dependent proliferation and (b) proliferation restricted to the bulk, away from the interface. Darker colours indicate later times.}
    \label{figS1}
\end{figure*}

To test the purported instability mechanism, we compare the density dependant growth regime at  $L = 640$, where we see the instability, with a growth regime where proliferation is restricted to the first 30 columns of lattice sites from the rear wall, but uniform within this region [Fig.\,\ref{figS1}]. We can see that introducing this regime with growth restricted to the bulk suppresses the instability, in agreement with our theory.
%%%%%%%%%%%%%%%%%%%%
%%%%%%%%%%%%%%%%%%%%%%%%
%%%%%%%%%%%%%%
%\bibliographystyle{apsrev} % To be used in case, we use revtex4 
%\bibliographystyle{apsrev4-1} % To be used in case, we use revtex4-1
\bibliography{references}
%%%%%%%%%%%%%%

\end{document}